\documentclass[11pt,letterpaper]{article}

\usepackage[utf8]{inputenc}
\usepackage[T1]{fontenc}
\usepackage{times}
\usepackage{graphicx}
\usepackage{amsmath,amssymb}
\usepackage{url}
\usepackage{hyperref}
\usepackage{booktabs}
\usepackage{multirow}
\usepackage{array}
\usepackage{caption}
\usepackage{subcaption}
\usepackage{placeins} 

\usepackage[margin=1in]{geometry}

\hypersetup{
    colorlinks=true,
    linkcolor=blue,
    citecolor=blue,
    urlcolor=blue,
}

\title{Beyond Prototyping: Autonomous, Enterprise-Grade Frontend Development from Pixel to Production via a Specialized Multi-Agent Framework}

\newcommand{\AuthorBlock}[3]{\parbox[t]{0.29\textwidth}{\centering \textbf{#1}\\#2\\\scriptsize #3}}
\author{\small  
\begin{tabular}{@{}p{0.30\textwidth}p{0.30\textwidth}p{0.30\textwidth}@{}}  
\AuthorBlock{Ramprasath Ganesaraja}{EdgeVerve Systems Limited}{ramprasath.g@edgeverve.com} &  
\AuthorBlock{Swathika N}{EdgeVerve Systems Limited}{swathika.n@edgeverve.com} &
\AuthorBlock{Saravanan AP}{EdgeVerve Systems Limited}{Saravanan\_P06@edgeverve.com} \\\\
\AuthorBlock{Kamalkumar Rathinasamy}{Infosys Limited}{Kamalkumar\_R@infosys.com} &
\AuthorBlock{Chetana Amancharla}{Infosys Limited}{Chetana\_Shanbhag@infosys.com} &
\AuthorBlock{Rahul Das}{EdgeVerve Systems Limited}{rahul.das05@edgeverve.com} \\\\
\AuthorBlock{Sahil Dilip Panse}{EdgeVerve Systems Limited}{SahilDilip\_Panse@edgeverve.com} &  
\AuthorBlock{Aditya Batwe}{EdgeVerve Systems Limited}{aditya.batwe@edgeverve.com} &
\AuthorBlock{Dileep Vijayan}{EdgeVerve Systems Limited}{Dileep\_Vijayan@edgeverve.com} \\\\  
\AuthorBlock{Veena Ashok}{EdgeVerve Systems Limited}{veena.ashok@edgeverve.com} &  
\AuthorBlock{Thanushree A P}{EdgeVerve Systems Limited}{thanushree.p02@edgeverve.com} & 
\AuthorBlock{Kausthubh J Rao}{EdgeVerve Systems Limited}{kausthubhj.rao@edgeverve.com} \\\\  
\AuthorBlock{Alden Olivero}{EdgeVerve Systems Limited}{alden.olivero@edgeverve.com} &  
\AuthorBlock{Roshan}{EdgeVerve Systems Limited}{roshan07@edgeverve.com} &  
\AuthorBlock{Rajeshwar Reddy Manthena}{Infosys Limited}{rajeshwar.manthena@infosys.com} \\\\   
\AuthorBlock{Asmitha Yuga Sre A}{Infosys Limited}{asmitha.a@infosys.com} &  
\AuthorBlock{Harsh Tripathi}{Infosys Limited}{harsh.tripathi01@infosys.com} & 
\AuthorBlock{Suganya Selvaraj}{Infosys Limited}{suganya.selvaraj04@infosys.com} \\\\
\AuthorBlock{Vito Chin}{Microsoft Corporation}{vito.chin@microsoft.com} &  
\AuthorBlock{Kasthuri Rangan Bhaskar}{BCT Digital}{krangan.ca@gmail.com} &  
\AuthorBlock{Binooj Purayath}{EdgeVerve Systems Limited}{Binooj\_Purayath@infosys.com} \\\\
\AuthorBlock{Venkatraman R}{EdgeVerve Systems Limited}{Venkatraman.R@edgeverve.com} &
\AuthorBlock{Sajit Vijayakumar}{EdgeVerve Systems Limited}{VSajit@infosys.com} &
\\  
\end{tabular}  
} 

\date{}

\begin{document}

\maketitle

\begin{abstract}
We present AI4UI, a framework of autonomous front-end development agents purpose-built to meet the rigorous requirements of enterprise-grade application delivery. Unlike general-purpose code assistants designed for rapid prototyping, AI4UI focuses on production readiness delivering secure, scalable, compliant, and maintainable UI code integrated seamlessly into enterprise workflows.
AI4UI operates with targeted human-in-the-loop involvement: at the design stage, developers embed a Gen-AI-friendly grammar into Figma prototypes to encode requirements for precise interpretation; and at the post processing stage, domain experts refine outputs for nuanced design adjustments, domain-specific optimizations, and compliance needs. Between these stages, AI4UI runs fully autonomously, converting designs into engineering-ready UI code.
Technical contributions include a Figma grammar for autonomous interpretation, domain-aware knowledge graphs, a secure abstract/package code integration strategy, expertise driven architecture templates, and a change-oriented workflow coordinated by specialized agent roles.
In large-scale benchmarks against industry baselines and leading competitor systems, AI4UI achieved 97.24\% platform compatibility, 87.10\% compilation success, 86.98\% security compliance, 78.00\% feature implementation success, 73.50\% code-review quality, and 73.36\% UI/UX consistency. In blind preference studies with 200 expert evaluators, AI4UI emerged as one of the leaders demonstrating strong competitive standing among leading solutions. Operating asynchronously, AI4UI generates thousands of validated UI screens in weeks rather than months, compressing delivery timelines while maintaining quality, maintainability, and strategic human oversight where it is most impactful.
\end{abstract}


\section{Introduction}

Enterprise front-end development faces a fundamental scalability challenge. Traditional workflows require extensive manual intervention at every stage—from design interpretation and code implementation to testing and deployment—creating bottlenecks that extend project timelines and increase costs. While recent advances in AI-assisted development tools have shown promise for rapid prototyping~\cite{gui2025webcode2m, si2025design2code}, they fall short of enterprise requirements where security, maintainability, compliance, and long-term scalability are non-negotiable.

The core problem lies in the gap between prototype-quality AI-generated code and production-ready enterprise applications. Current solutions typically focus on visual fidelity—converting designs to functional interfaces—but neglect the comprehensive infrastructure required for enterprise deployment: authentication systems, validation frameworks, performance optimization, security compliance, and integration with existing enterprise architectures~\cite{xiao2025designbench}.

This paper introduces AI4UI, a framework of autonomous front-end agents specifically engineered for enterprise-grade application development. Prior multi-agent and prototype-focused systems emphasize assisted or confirmation-based generation workflows rather than fully autonomous, production-oriented delivery \cite{ding2025frontenddiffusion, xiao2024prototype2code, yuan2024maxprototyper, yuan2024prototypeagent, honarvar2024automatic}. Surveyed Web-automation and agent frameworks highlight interface-level task solving but stop short of end-to-end enterprise integration \cite{ning2025webagents, ikumapayi2023automated}. Model-driven and safety-oriented UI efforts address abstraction, multiexperience, and user agency without establishing a continuous autonomous pipeline to production \cite{planas2021modeldriven, lu2025designingui, thomsy2025genie}.

Our key contributions are:

\begin{enumerate}
\item A novel LLM-friendly grammar embedded within Figma that enables comprehensive design specification capture, including business logic, validation rules, and implementation constraints.

\item A domain-aware knowledge graph architecture that maintains contextual relationships across large-scale applications, ensuring consistent component reuse and architectural integrity.

\item A secure package abstraction methodology that enables integration of proprietary enterprise functionality while maintaining code generation consistency.

\item An expertise-driven agent architecture that leverages industry best practices and established library patterns to generate maintainable, production-quality code.

\item Comprehensive benchmarking against industry leaders across multiple dimensions including compilation success, feature implementation, code review quality, and security compliance.
\end{enumerate}

The evaluation demonstrates that AI4UI achieves competitive performance with leading commercial solutions while providing the enterprise-grade capabilities required for production deployment. In expert preference studies with 200 evaluators, AI4UI emerged as one of the leaders, validating both its technical capabilities and practical applicability.

The remainder of this paper is organized as follows: Section~\ref{sec:related_work} reviews related work in AI-driven frontend development; Section~\ref{sec:architecture} details the AI4UI system architecture and technical innovations; Section~\ref{sec:benchmarking} describes our benchmarking methodology and evaluation framework; Section~\ref{sec:results} presents comprehensive results across multiple performance dimensions; and Section~\ref{sec:future},~\ref{sec:conclusion} discusses implications and future directions.

\section{Related Work}
\label{sec:related_work}
Recent years have seen rapid progress in automating front-end development using artificial intelligence, particularly in the conversion of design artifacts into production-ready code. Several frameworks and benchmarking efforts have emerged to address the challenges of code quality, maintainability, and enterprise readiness.

\subsection{Traditional Front-End Development Workflows}

The traditional process begins with the design stage, where UI/UX specifications are created entirely by designers through iterative discussions and manual refinement. These designs are then translated into implementation plans by developers, who manually code, integrate, and configure components. Throughout development, engineers remain actively engaged in handling state management, validations, performance tuning, and feature adjustments. In the final stage, quality assurance teams perform extensive manual testing and refinements before deployment, ensuring the application meets all functional and compliance requirements~\cite{garrett2010elements, nielsen2012usability, sharp2019interaction}.

\begin{figure}[!htbp]
\centering
\includegraphics[width=0.85\textwidth]{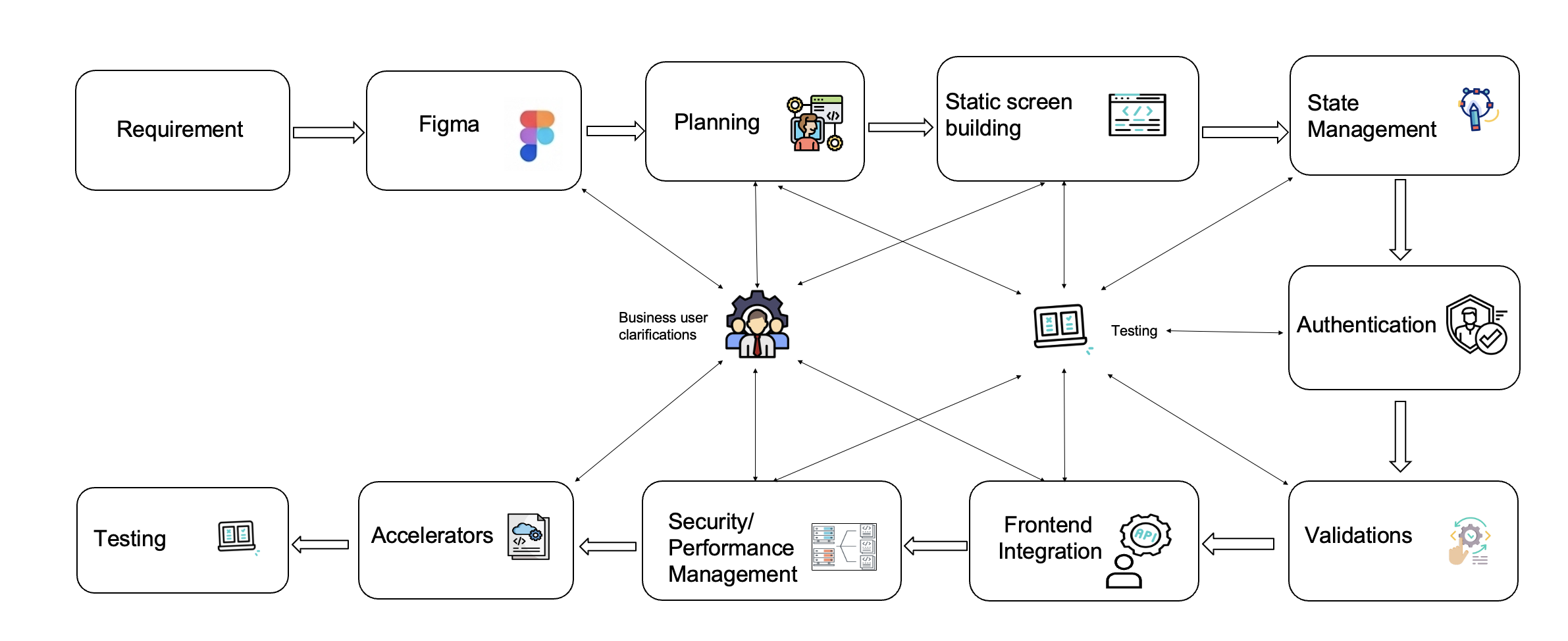}
\caption{Traditional enterprise front-end development workflow showing manual, iterative processes from requirements to deployment.}
\label{fig:ai4ui_architecture}
\end{figure}
\FloatBarrier

\subsection{AI-Driven Design-to-Code Systems}

WebCode2M proposes a multi-modal approach for generating web code from design inputs such as images and textual descriptions. The system benchmarks code generation quality and highlights the gap between rapid prototyping and production-grade code, which is critical for enterprise applications~\cite{gui2025webcode2m}.

Design2Code presents an end-to-end pipeline for converting Figma designs into React code. It emphasizes semantic understanding of design intent and code maintainability, and incorporates human-in-the-loop refinement stages to improve output quality~\cite{si2025design2code}.

DesignBench introduces a large-scale benchmark for evaluating AI systems that generate frontend code from design inputs. It covers metrics such as code correctness, UI/UX consistency, and compilation success, providing a standardized evaluation framework for design-to-code systems~\cite{xiao2025designbench}.

\subsection{Multi-Agent Front-End Systems}
Recent multi-agent front-end efforts focus on assisted prototyping, staged confirmation, or design-to-code translation rather than uninterrupted autonomous enterprise delivery. Frontend Diffusion introduces agentic workflows for self-representative code generation but does not address enterprise compliance layers \cite{ding2025frontenddiffusion}. Prototype2Code constructs layout trees and generates functional code from UI prototypes with limited operational hardening \cite{xiao2024prototype2code}. MAxPrototyper and PrototypeAgent concentrate on automatic wireframing, interactive prototype generation, and clarifying implicit design intent through specialized sub-agents \cite{yuan2024maxprototyper, yuan2024prototypeagent}. Figma-centric plugin and augmentation approaches enable assisted UI creation but remain designer-driven \cite{honarvar2024automatic}. Automated OpenAI-based code generation identifies logical UI parts yet lacks multi-layer enterprise validation \cite{ikumapayi2023automated}. Genie explores context-aware multi-agent customization of home-screen interfaces without full production workflow integration \cite{thomsy2025genie}. WebAgents surveys GUI-driven task-solving agents, framing broader automation directions rather than production-grade front-end engineering \cite{ning2025webagents}. Model-driven multiexperience UI generation and safety-oriented UI/agent workflows emphasize specification abstraction and user control, not autonomous lifecycle execution \cite{planas2021modeldriven, lu2025designingui}. Our work differs by coupling design grammar, domain knowledge graphs, change-oriented orchestration, and compilation integrity into a single autonomous pipeline targeting enterprise readiness.

\subsection{Commercial Tools and Benchmarks}

Kombai is a commercial tool that uses AI to convert Figma designs to React code. Benchmarking reports from Kombai compare its performance against other tools on metrics including code review quality, feature implementation, and compilation success~\cite{kombai2024}.

Design Arena focuses on community-driven evaluation of AI-generated UI designs through pairwise comparisons, where rankings emerge from collective preferences rather than curated opinions. This approach measures user preference and visual fidelity through direct, anonymized comparisons~\cite{designarena2024}.

\subsection{Generalist and Specialist AI Agents}

AGIEval benchmarks both generalist and specialist AI agents on complex tasks, including software engineering. The findings support the use of specialized agents for improved performance in enterprise and technical domains~\cite{zhong2023agieval}.

The reviewed works demonstrate a growing interest in automating frontend engineering with AI, particularly in converting design artifacts to code, ensuring production readiness, and benchmarking AI-generated outputs. These efforts collectively advance the field toward robust, scalable, and maintainable UI development.

\section{Autonomous Agent Architecture with Strategic Human Oversight}
\label{sec:architecture}
Rather than replacing developers, our autonomous front-end agents are designed to amplify human expertise at strategic touchpoints. Human involvement occurs at two key stages:
\begin{itemize}
\item \textbf{Design Stage:} Developers and designers integrate the LLM-friendly grammar into Figma prototypes, encoding requirements, constraints, and behaviors for accurate agent interpretation.
\item \textbf{Post-Processing Stage:} After autonomous generation, domain experts refine outputs to address nuanced design interpretations, domain-specific optimizations, or additional compliance adjustments.
\end{itemize}
Between these stages, AI4UI operates in fully autonomous mode, translating design inputs into production ready UI screens complete with state management, validation logic, and integration capabilities. This workflow enables the asynchronous generation of thousands of high-fidelity, engineering-ready screens within weeks, accelerating deployment while maintaining quality through focused human oversight.

\subsection{Multi-Agent Workflow Architecture}

\begin{figure}[!htbp]
\centering
\includegraphics[width=0.85\textwidth]{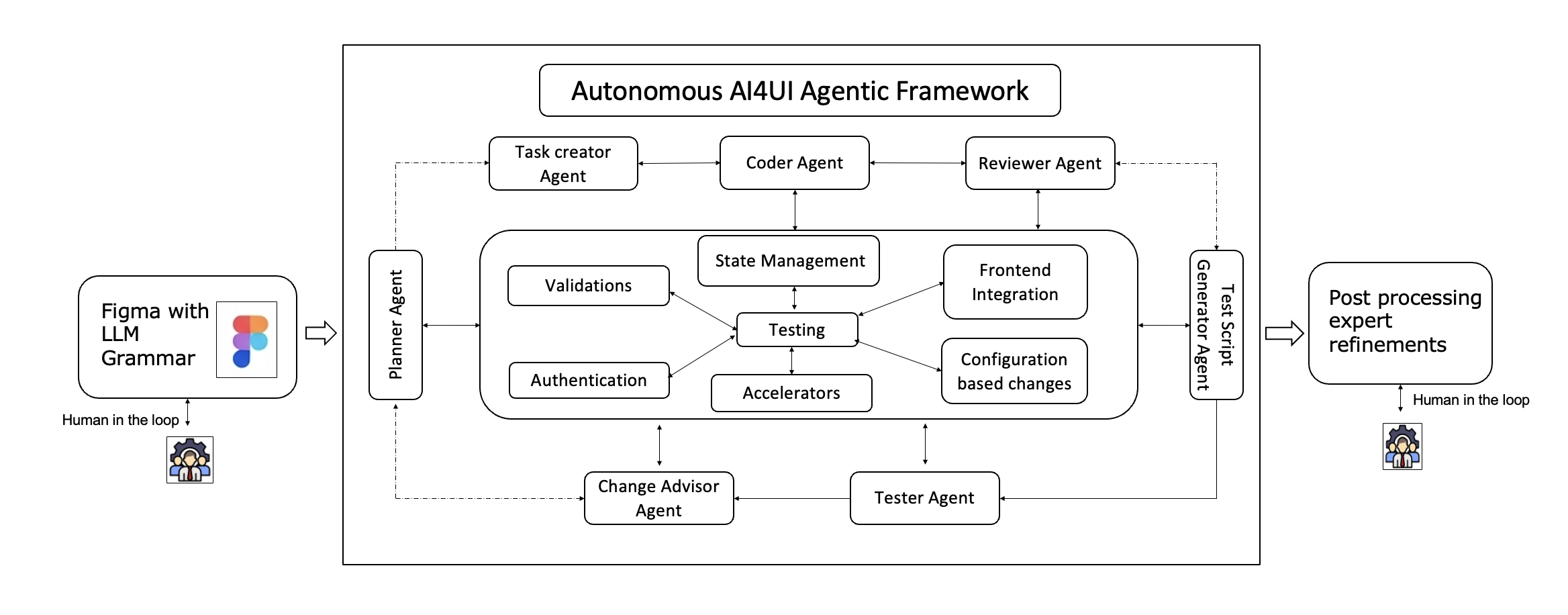}
\caption{AI4UI autonomous agent workflow showing specialized agents coordinating design interpretation, planning, coding, review, testing, and change management.}
\label{fig:architecture_workflow}
\end{figure}
\FloatBarrier

\subsection{Enterprise-Grade Capabilities}

\subsubsection{Core Enterprise Requirements}

Enterprise applications must satisfy architectural quality attributes such as scalability, performance efficiency, security, maintainability, reliability, portability, and interoperability~\cite{iso25010,bass2012software}. Security hardening and systematic vulnerability reduction require adherence to established verification standards and secure design principles~\cite{owaspasvs,mcgraw2006software}. Modern integration and evolutionary scalability further depend on service decomposition and loosely coupled interface strategies~\cite{newman2015microservices}. Complementing these engineering foundations, user trust and perceived credibility remain influenced by visual clarity and aesthetic quality~\cite{fogg2001what, robins2008aesthetics}. AI4UI's agents are engineered for continuous, sustained operation over extended periods without manual intervention.
\begin{itemize}
\item \textbf{Authentication Systems:} Robust authentication mechanisms ensure that only authorized users can access specific application areas. The platform supports layered user verification and role-based permissions, protecting sensitive operations and data while maintaining a secure user experience.

\item \textbf{Validation Frameworks:} All user inputs are checked for accuracy, completeness, and format compliance through a combination of automated rules and configurable checks. This reduces errors, prevents invalid data from entering the system, and ensures reliable application behaviour.

\item \textbf{Security Implementation:} Security is embedded in every aspect of the application design, from data handling to communication between components. Protective measures safeguard against vulnerabilities, maintain data confidentiality, and support organizational risk management.

\item \textbf{Performance Optimization:} The system is designed to remain responsive and reliable under varying workloads. Resource usage is managed efficiently, and application behaviour is tuned to deliver consistent performance for both small and large user bases.

\item \textbf{Platform Compatibility:} The application operates consistently across multiple browsers, devices, and operating environments. Layouts and behaviours adapt smoothly to different screen sizes and capabilities, ensuring accessibility for a wide range of users.

\item \textbf{UI Consistency:} A unified design approach ensures that all screens and components share a common visual language, structure, and interaction style. This improves usability and reinforces the product's identity.

\item \textbf{Extensibility:} The platform uses an adaptable architecture that can evolve over time. New features or integrations can be added without disrupting existing functionality, supporting long-term growth and changing business needs.

\item \textbf{Traceability:} All changes to the application, from configuration updates to content modifications are recorded and can be reviewed later. This history supports accountability, compliance needs, and easier issue resolution.
\end{itemize}

\subsubsection{Operational Enterprise Features}

\begin{itemize}
\item \textbf{Rapid Integration:} Our autonomous front-end agents enable rapid alignment with Backend-for-Frontend (BFF) API layers during early development stages by connecting exclusively through stub values rather than the actual backend. This stub-based integration approach allows teams to validate UI flows, API contracts, and data handling logic without depending on live backend availability or stability. Stub responses are generated to match real-world payload structures, ensuring the front-end can be tested for correctness, performance, and error handling in isolation.

\item \textbf{Configuration-Based Changes:} To support global scalability and adaptability, non-functional updates such as multilingual enablement, label adjustments, and text replacements are driven entirely through configuration files, not through invasive code changes. This approach centralizes customization logic, reducing regression risks and improving maintainability.

\item \textbf{Agile Development Cycle:} Our autonomous front-end agents support highly dynamic development workflows by enabling rapid iteration and deployment of UI components. Changes can be implemented and verified quickly through generated application screens, allowing teams to assess functionality, design consistency, and user experience after each build cycle. By streamlining the path from design to deployment, the system adheres to agile methodology principles while ensuring enterprise-grade robustness and quality control.

\item \textbf{State Management:} Complex applications often require handling multifaceted states ranging from asynchronous data loads to intricate conditional rendering. Our agents incorporate advanced state management capabilities that synchronize front-end components with underlying business processes in real time. Our solution automatically tracks user interactions, session data, and application state transitions to prevent inconsistencies, stale UI, or unintended behavior.

\item \textbf{Accelerators:} We provide a suite of configurable accelerators that can be enabled per project. These accelerators are implemented as extensible, auditable components so teams can enforce privacy, formatting, and integration rules without changing core application logic.

\item \textbf{Validations:} We implement a combinational approach to UI validation that merges intelligent, schema-free type inference with user-defined customization. The system autonomously analyzes each field's name and contextual usage to accurately detect data types such as email, phone number, date, numeric, and currency and then apply lightweight, auditable type-specific checks without requiring manual code for every validator. In addition to this automated logic, users can introduce custom validation rules to handle domain-specific or edge-case requirements, ensuring both flexibility and precision. This hybrid model reduces development effort while allowing teams to maintain strict quality and compliance standards across diverse application contexts.
\end{itemize}

\subsection{Technical Innovations}

While prior work focuses primarily on mapping designs to code and on establishing benchmarks for correctness and visual fidelity~\cite{gui2025webcode2m, si2025design2code, xiao2025designbench}, AI4UI advances the state of the art in several distinct ways that target enterprise requirements beyond pixel fidelity.

\subsubsection{Core Technical Contributions}

\begin{itemize}
\item \textbf{LLM-Friendly Grammar for Figma:} Figma is designed for interactive prototypes, design systems, and real-time collaboration. It helps developers understand user flows and implement interfaces, but it was not created for autonomous agents. We built a Gen AI-first grammar inside Figma to capture every user requirement and all relevant implementation details so agents can interpret designs reliably and act on them. The grammar encodes intent, screen states, transitions, validation rules, edge cases, accessibility notes, business constraints, asset mappings, and testing criteria. For example, when a designer creates multiple states of the same screen, the grammar differentiates each variant in the prototype so agents can identify, interpret, and handle them accurately.

\item \textbf{Fit-for-Purpose Knowledge Graphs:} Embedding enables the creation of code snippets and simplifies search, which works well for text-based content. However, for front-end code, understanding domain-specific code relationships is essential for successfully implementing features and ensuring high-impact code quality. For instance, in large-scale applications with hundreds of generated screens, the contextual linkage between prop values and reusable components can easily be lost without proper relationship tracking. To address this, we develop domain-aware knowledge graphs that make it straightforward to retrieve both the required information and its associated context across the entire enterprise application.

\item \textbf{Abstract and Package Approach:} In enterprise environments, maintaining code consistency is a critical requirement. Ensuring that central pipeline code generation remains uniform across the organization is essential for scalability, maintainability, and quality control. At the same time, enterprises often need to integrate their internal code libraries, which encapsulate proprietary functionalities and security logic. These internal implementations are confidential and not publicly accessible yet must be incorporated seamlessly into enterprise systems. To address this need, we propose an approach in which AI agents first generate a standalone package, abstracted from the central pipeline execution flow. The methodology achieves a balance between consistent, reproducible code generation and the flexibility to embed proprietary functionality securely within enterprise applications.

\item \textbf{Expertise-Driven Architecture:} Shipping production-quality code with AI agents is possible, but only under one caveat: we need to know what `good' looks like. Our expertise-driven architecture ensures we always ship the best quality code, not just code that compiles, but code that remains robust and maintainable over time. By leveraging frontend experts' guidance and popular library combinations across versions and having these approaches reviewed by industry-leading frontend builders, we make sure AI agents amplify expertise rather than create technical debt.
\end{itemize}

\subsubsection{Advanced Workflow Innovations}

\begin{itemize}
\item \textbf{Multiple Variation Exploration:} Even with expertise, experimentation remains crucial for finding optimal solutions. Our autonomous agents experiment with multiple variations in parallel. For example, testing React DnD, DnD kit, and React Beautiful DnD for drag-and-drop components and then selecting the best fit for the given context and scale requirements.

\item \textbf{Change-Oriented Development Workflow:} In enterprise front-end development, ad-hoc code modifications often lead to inconsistencies, regressions, and skipped requirements. Our approach addresses this by first capturing every planned UI change as a detailed Request for Change (RFC), documenting scope, impacted components, dependencies, and acceptance criteria. These RFCs are then processed through a coordinated suite of autonomous agents which includes a Designer for UI/UX specifications, an Orchestrator for sequencing tasks, a Planner for implementation strategies, a Reviewer for quality checks, and an Implementer for code delivery. This architecture enables teams to identify which tasks should be executed sequentially and which can run in parallel, striking the right balance between code quality and performance. Also, the structured workflow ensures that required updates are implemented fully, consistently, and without unintended side effects, resulting in clean, production-ready code.

\item \textbf{Context-Aware State Management:} Autonomous agents can sometimes miss complex or rare state transitions, especially in large applications with dynamic user flows. We solve this by integrating knowledge graphs, the Figma grammar, and the RFC process into a unified state management system. This ensures that state handling logic covers not only common cases, but also rare, high-impact scenarios that could otherwise slip through. The result is a highly reliable treatment of component lifecycle, user interactions, and error states that are crucial for enterprise UX stability.

\item \textbf{Layered Compilation Integrity Module:} In automated code generation for enterprise applications, build errors can cascade: small syntax issues may mask larger architectural problems. Our compilation process applies a layered, bird's-eye strategy to ensure complete integrity before deployment. First, autonomous agents address the most localized, granular errors (such as minor syntax or missing imports), then systematically progress to higher-level issues like component interdependencies or build tooling misconfigurations. Between each correction stage, functionality checks are run to detect any inadvertent alteration of existing business logic, and corrective measures are applied immediately if deviations are found. This staged approach, coordinated by Orchestrator, Planner, Reviewer, and Implementer agents ensures not only a clean codebase that compiles successfully, but also preserves intended application behavior across the entire generated system.
\end{itemize}

Together, these contributions shift the research focus from isolated design-to-code mappings toward a systems view that addresses enterprise concerns: privacy, traceability, maintainability, and predictable integration. In contrast to benchmarks that emphasize isolated metrics, AI4UI couples evaluation with workflow primitives (RFCs, knowledge graphs, and package abstractions) so that generated outputs are not only correct and visually consistent, but also deployable and auditable within enterprise software lifecycles.

\section{Benchmarking and Performance Evaluation}
\label{sec:benchmarking}
We conducted comprehensive state-of-the-art testing across multiple critical metrics to validate our agents' performance against industry standards and competing solutions~\cite{zheng2023judging, chiang2024chatbot}. The evaluation followed a structured, multi-phase methodology to ensure accuracy, reproducibility, and fairness.

To obtain an objective and multi perspective assessment, we employed both large language model (LLM)–based evaluation and expert human review~\cite{dubois2024length}. The LLM served as an automated evaluator, applying consistent criteria across all test cases, while human evaluators provided qualitative and contextual insights~\cite{amabile1982social, amabile1996creativity}. The final performance score for each metric was calculated as the average of the LLM generated score and the human assigned score, balancing computational objectivity with human judgment and domain expertise.

\subsection{Test Environment Setup}

All benchmarks were run in controlled, reproducible environments matching typical enterprise development setups. This included standardized build tooling, dependency versions, and network conditions to ensure that differences in results stemmed from agent capabilities rather than environmental factors.

\subsection{Dataset and Task Selection}

We assembled a representative set of UI development tasks covering varied complexity: from simple component creation to multi-screen flows with advanced state transitions, validation rules, and responsive layouts. These tasks were designed to reflect real-world enterprise requirements and mapped directly to the evaluation dimensions (Compilation Success, Code Review, Feature Implementation, etc.).

\subsection{Agent Execution and Logging}

Each task was executed end-to-end by the autonomous front-end agents. Detailed logs captured execution steps, intermediate decisions, and the final generated code.

\subsection{Objective Metric Calculation}

For each dimension, objective scoring criteria were applied:

\begin{itemize}
\item \textbf{Code Review:} Code maintainability, clarity, adherence to naming conventions, cleanliness, and robustness in error handling were assessed through cyclomatic complexity checks, style guide compliance, documentation frequency analysis, and error-handling coverage. This evaluation was supported by static analysis tools such as ESLint, alongside formatting checks performed through Prettier to ensure code consistency and readability.

\item \textbf{Feature Implementation:} Implementation completeness and functional accuracy were validated against the specification, ensuring correct imports, modularity, and logical structure. Unit and integration test coverage was measured using Jest, while import structure and modular design were analyzed via static inspection tools.

\item \textbf{Compilation Success:} Build reliability was evaluated by tracking success rates, warning/error counts, dependency integrity, and deterministic output consistency~\cite{chen2021evaluating}. TypeScript compiler checks validated type safety, while npm and yarn audits were used to ensure dependency health. Key CI/CD build logs were examined for reproducibility and consistent build outputs, verifying the stability of the build process across multiple runs.

\item \textbf{Performance:} Application efficiency was measured across three key dimensions: Time (component render and commit times), Fluidity (FPS counters and frame drop detection), and Efficiency (identification of unnecessary re-renders or oversized component trees). These metrics were obtained using React DevTools and the React Native Performance Monitor, providing a quantitative assessment of responsiveness and rendering optimization.

\item \textbf{Platform Compatibility:} Cross-platform consistency was tested to ensure parity between iOS and Android environments, verifying feature behavior, layout accuracy, and rendering fidelity.

\item \textbf{Security:} Security evaluation included static analysis for secret leaks and vulnerabilities, dependency scanning, and rigorous input validation reviews. ESLint security plugins and npm audit tools were employed to detect potential risks. Logging policies and data handling practices were also verified to comply with secure coding guidelines.

\item \textbf{UI/UX Consistency:} User interface and experience consistency were assessed for alignment with the established design system, cohesive styling, and component reuse~\cite{dou2019webthetics, lu2014rapid, vonwangenheim2018agree}.
\end{itemize}

\subsection{User Preference}

While Design Arena relies on community-driven pairwise comparisons to evaluate the visual quality of AI-generated designs~\cite{designarena2024}, its current leaderboard highlights Flames.blue, Mocha, and Magic Patterns as the top performers.

\begin{itemize}
\item \textbf{Side-by-Side Reviews:} Participants reviewed generated interface designs from AI4UI and competitor tools under identical task prompts.

\item \textbf{Blind Comparison:} Designs were presented without branding or tool identifiers to remove bias.

\item \textbf{Scoring \& Selection:} Reviewers selected the design they would prefer to use or deliver to production, considering both visual attractiveness and functional clarity.

\item \textbf{Evaluator Mix:} The evaluation group included professional designers, front-end engineers, and representative end-users to ensure results reflected both technical feasibility and real-world appeal~\cite{reinecke2013predicting, pollitt2012comparative, whitehouse2012using}.
\end{itemize}

\subsection{Comparative Analysis}

Scores from our agent runs were compared against publicly reported figures~\cite{kombai2024} for available competitor systems (e.g., Kombai, Sonnet + Agent, Gemini + Agent). Where public data was unavailable, the values recorded here serve as internal baselines for future benchmarks.

\section{Results}
\label{sec:results}
AI4UI demonstrates strong engineering readiness across multiple evaluation dimensions. Headline outcomes include 78.0\% feature implementation, 73.5\% code-review quality, and 87.1\% compilation success; AI4UI also reports 97.24\% platform compatibility and 86.98\% security. These results indicate that AI4UI not only reproduces visual designs but also generates engineering-ready code at scale.
\begin{table}[!htbp]
\centering
\caption{Performance evaluation: comparison of AI4UI against industry averages and a selected benchmark.}
\label{tab:performance_comparison}
\small
\begin{tabular}{p{3.5cm}ccc>{\raggedright\arraybackslash}p{4cm}}
\toprule
\textbf{Evaluation Category} & \textbf{Industry Average} & \textbf{Industry Leader} & \textbf{Our Performance} & \textbf{Performance Comments w.r.t. Industry Leader} \\
\midrule
Code Review & 42.25\% & 72\% & 73.5\% & +1.5\% Modest Uplift \\
Feature Implementation & 25\% & 43\% & 78\% & +35\% Substantial Improvement \\
Compilation Success & 60.25\% & 96\% & 87.1\% & -8.9\% Gap, Second Highest Performer \\
UI/UX Consistency & N/A & N/A & 73.36\% & Baseline metric for future benchmarking \\
Platform Compatibility & N/A & N/A & 97.24\% & Baseline metric for future benchmarking \\
Performance & N/A & N/A & 72.06\% & Baseline metric for future benchmarking \\
Security & N/A & N/A & 86.98\% & Baseline metric for future benchmarking \\
\bottomrule
\end{tabular}
\end{table}
\FloatBarrier
*Additional tools used in our workflow include Locofy MCP and MCP Context7, which, together with our AI4UI agents, enable the automated conversion of Figma designs into production-ready React and React Native code.

The results presented here reflect the direct, unmodified output generated by the AI agents during evaluation. In practical deployment, an additional human-augmentation stage is applied, wherein domain experts perform targeted manual refinements. This post-processing significantly enhances quality, and the resultant final product typically achieves performance metrics exceeding those reported in this baseline.

\subsection{Per-metric Comparisons}
We present three focused comparisons that emphasize engineering readiness: Code Review, Feature Implementation, and Compilation Success.

The figure below presents a comparative analysis of performance across three key evaluation categories: Code Review, Feature Implementation, and Compilation Success. The results are shown for three groups: the industry benchmark (representing the average performance of commonly used tools), the industry leader (Kombai), and AI4UI. In this context, "Industry Leader" refers to Kombai, while "Industry Benchmark" aggregates other evaluated models and platforms. AI4UI demonstrates substantial improvements over the industry benchmark in all categories, with particularly strong results in feature implementation and code review. Although Kombai achieves the highest compilation success rate, AI4UI closely follows, outperforming other benchmarked solutions. This comparison highlights AI4UI's engineering readiness and its competitive standing relative to both the market leader and the broader landscape of AI-assisted front-end development tools.
\begin{figure}[!htbp]
\centering
\includegraphics[width=0.75\textwidth]{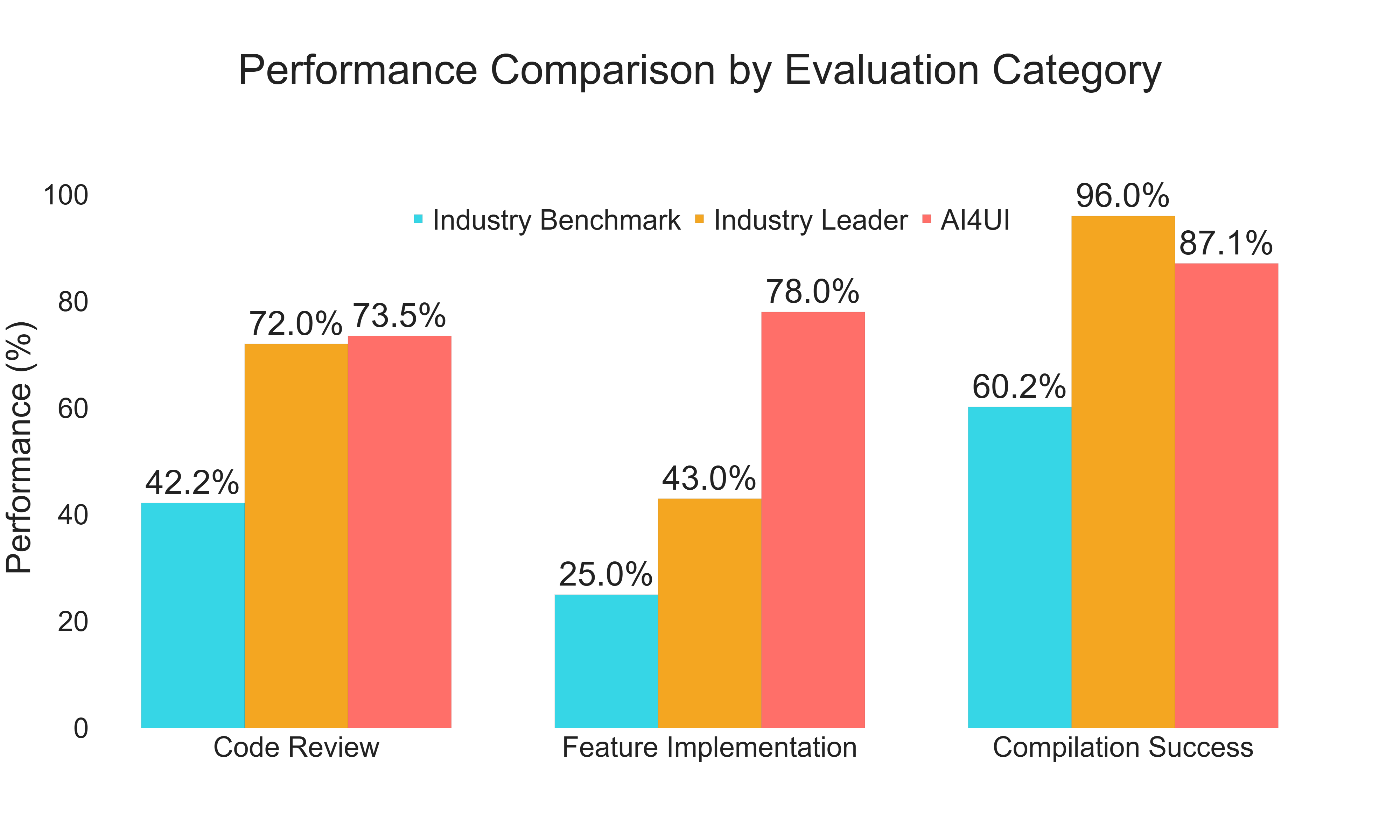}
\caption{Bar chart comparing AI4UI and industry benchmark performance across code review, feature implementation, and compilation success.}
\label{fig:bar_comparison}
\end{figure}
\FloatBarrier

\subsubsection{Code Review}
\begin{figure}[!htbp]
\centering
\includegraphics[width=0.85\textwidth]{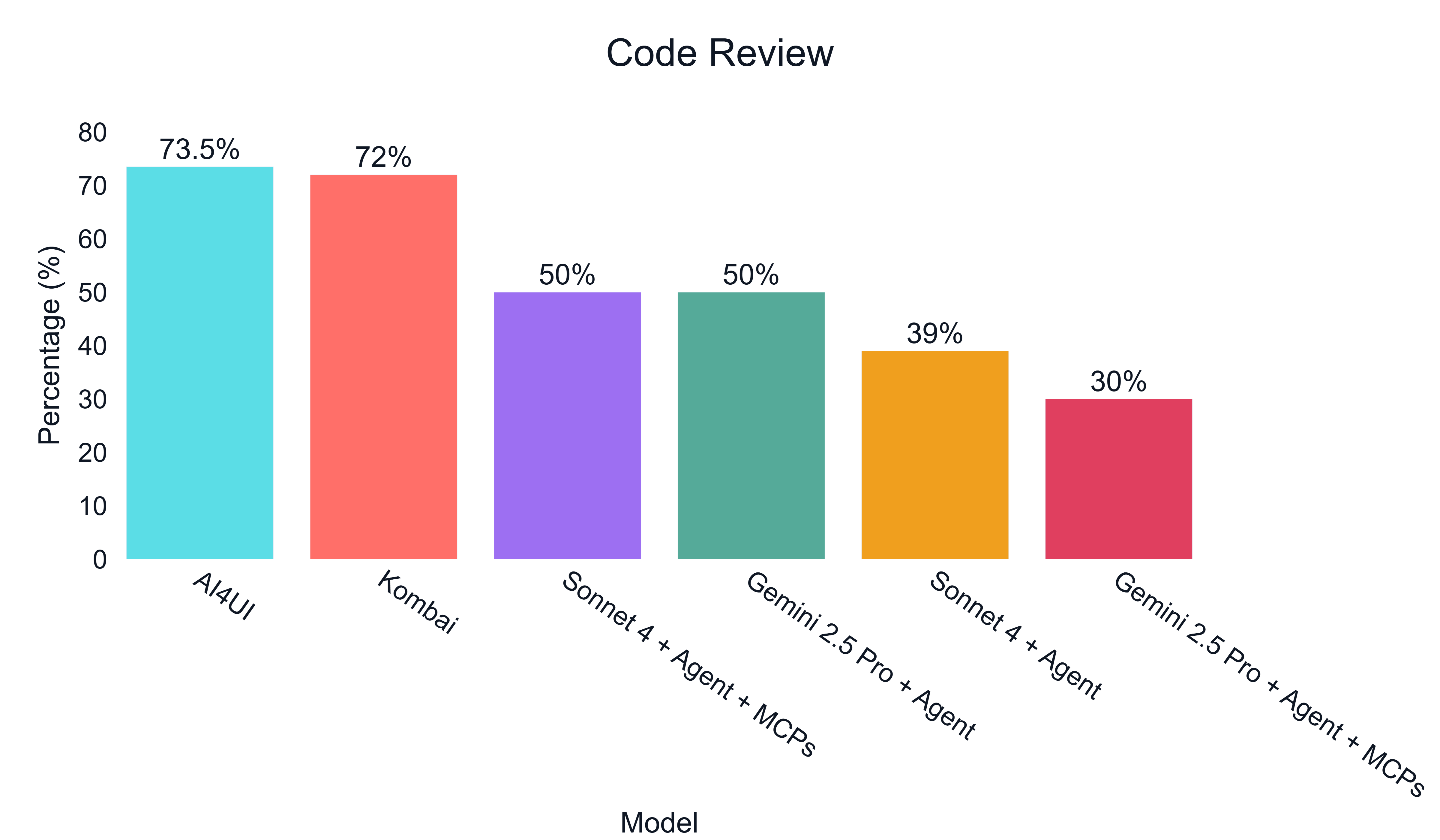}
\caption{Code review accuracy comparison across models, highlighting AI4UI as the leader.}
\label{fig:code_review}
\end{figure}

AI4UI achieves 73.5\% on the code-review metric slightly above the industry leader benchmark (72\%) indicating generated code adheres closely to maintainability and naming conventions, reducing expected manual cleanup (Figure~\ref{fig:code_review}).

Code review accuracy is a critical metric for assessing the maintainability, readability, and overall quality of automatically generated code. It reflects how well the output adheres to established coding standards, naming conventions, and best practices, thereby minimizing the need for manual intervention and post-generation cleanup. The comparative analysis presented below evaluates multiple autonomous front-end development models on this metric. AI4UI achieves the highest code review score at 73.5\%, marginally outperforming the industry leader Kombai (72\%). Other benchmarked models, including Sonnet 4 + Agent, Gemini 2.5 Pro + Agent, and their MCP-augmented variants, demonstrate significantly lower code review accuracy, with scores ranging from 30\% to 50\%. These results highlight AI4UI’s effectiveness in producing clean, maintainable code that aligns with enterprise engineering standards and reduces developer effort in the review process.
\FloatBarrier

\subsubsection{Feature Implementation}
\begin{figure}[!htbp]
\centering
\includegraphics[width=0.85\textwidth]{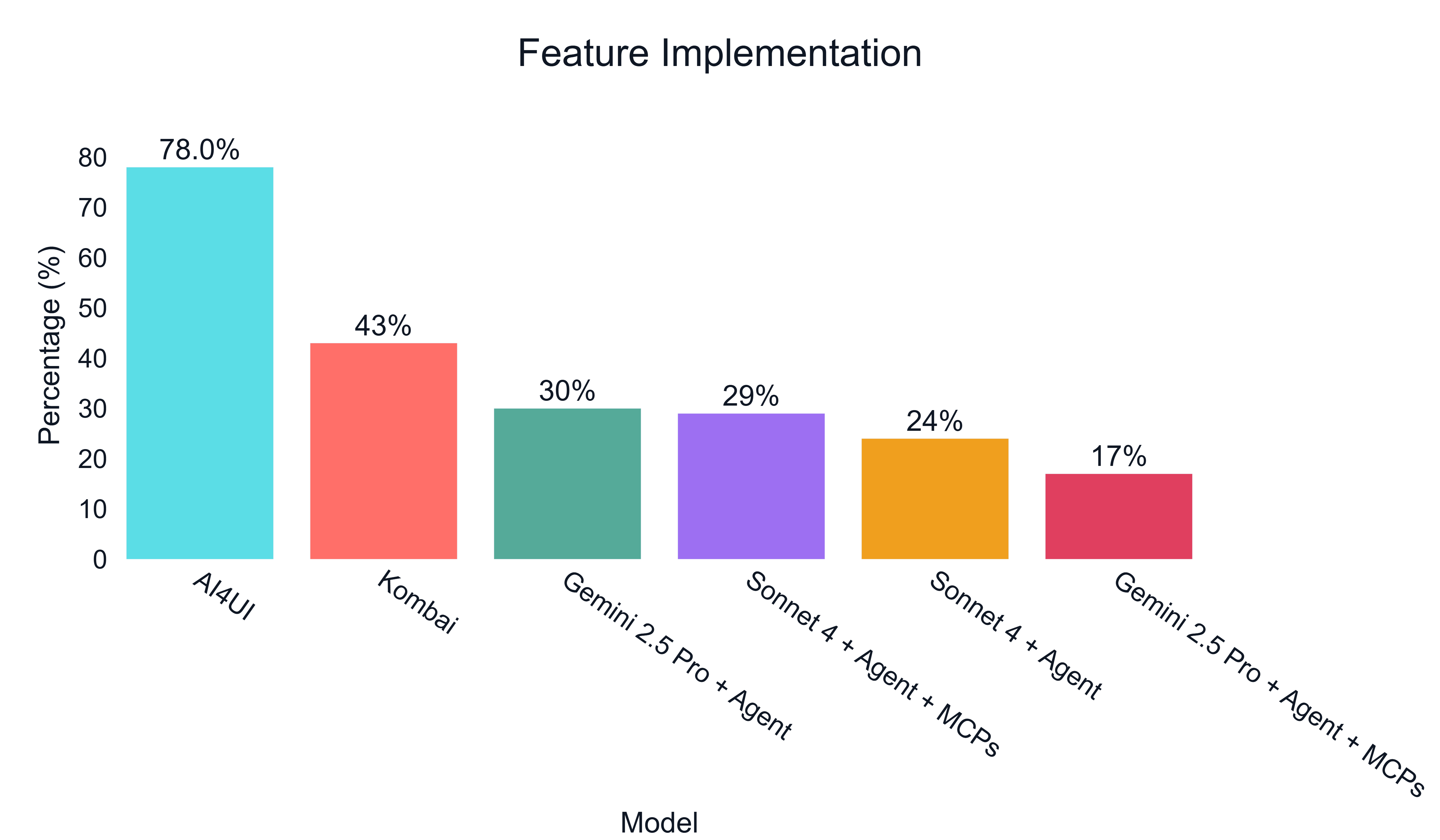}
\caption{Feature implementation rates across models, with AI4UI leading at 78\%.}
\label{fig:feature_impl}
\end{figure}

With a 78.0\% feature implementation rate, AI4UI considerably outperforms the internal benchmark (25\%) and an industry leader (43\%), implying large reductions in developer effort required to finalize features produced from designs (Figure~\ref{fig:feature_impl}).

Feature implementation rate is a key indicator of how effectively an autonomous front-end development system translates design specifications into functional code components. It measures the proportion of features from the original design that are correctly and fully realized in the generated output, minimizing the need for manual completion or correction by developers. The comparative analysis below demonstrates that AI4UI achieves a feature implementation rate of 78.0\%, substantially outperforming both the industry leader Kombai (43\%) and other benchmarked models, which range from 17\% to 30\%. This significant margin highlights AI4UI’s capability to autonomously deliver complete, production-ready features, thereby reducing developer workload and accelerating the overall development cycle in enterprise environments.
\FloatBarrier

\subsubsection{Compilation Success}
\begin{figure}[!htbp]
\centering
\includegraphics[width=0.85\textwidth]{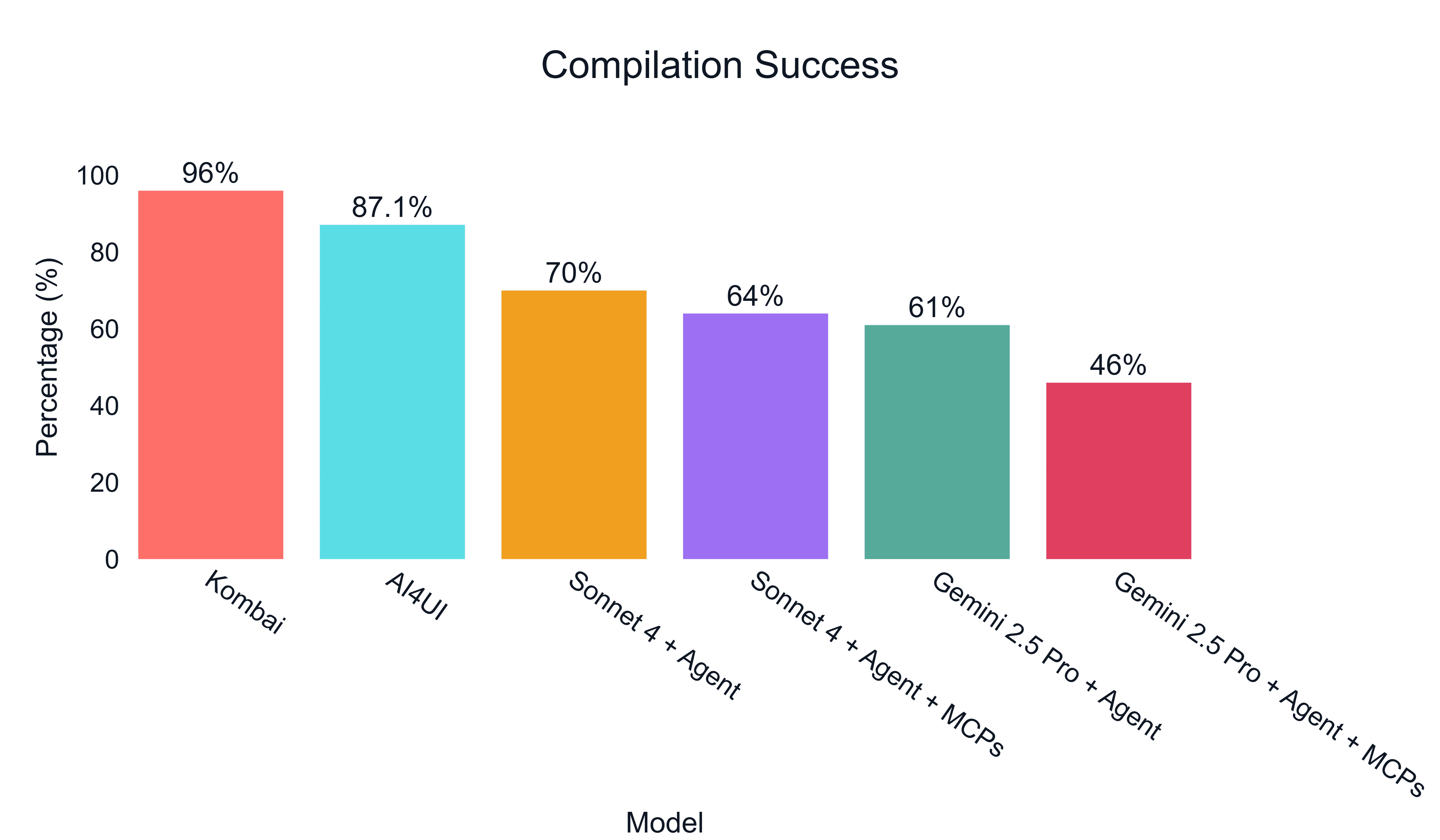}
\caption{Compilation success comparison, showing Kombai highest and AI4UI at 87.1\%.}
\label{fig:compilation}
\end{figure}

AI4UI achieves an 87.1\% compilation success rate, demonstrating robust build reliability; our layered compilation integrity process further reduces developer time spent resolving build issues and improves deployment readiness (Figure~\ref{fig:compilation}).

Compilation success is a fundamental metric for evaluating the reliability and production readiness of autonomous front-end development systems. It measures the proportion of generated code that builds successfully without errors, reflecting the system’s ability to produce deployable outputs that meet enterprise standards. The comparative analysis below presents compilation success rates for leading platforms and models. Kombai, the industry leader, achieves the highest rate at 96\%, while AI4UI demonstrates robust performance with an 87.1\% success rate. Other benchmarked models, including Sonnet 4 + Agent, Gemini 2.5 Pro + Agent, and their MCP-augmented variants, exhibit lower compilation success rates ranging from 46\% to 70\%. These results underscore AI4UI’s strong engineering reliability and its competitive standing relative to both the market leader and the broader landscape of AI-assisted front-end generation tools.
The radar chart below provides a comprehensive visualization of AI4UI’s performance across six critical quality dimensions: correctness, code quality, performance, compatibility, UI/UX, and security. Each axis represents a distinct evaluation criterion, with the plotted values indicating the percentage score achieved in each area. AI4UI demonstrates exceptional platform compatibility (97.2\%) and robust security compliance (87.0\%), alongside strong results in correctness (77.5\%), code quality (73.5\%), UI/UX consistency (73.4\%), and performance (72.1\%). This balanced distribution of scores highlights the system’s ability to deliver enterprise-grade front-end solutions that are secure, compatible across platforms, maintainable, performant, and visually consistent. The multidimensional assessment underscores AI4UI’s readiness for production deployment in demanding enterprise environments, where holistic quality across all dimensions is essential.

\FloatBarrier
\begin{figure}[!htbp]
\centering
\includegraphics[width=0.7\textwidth]{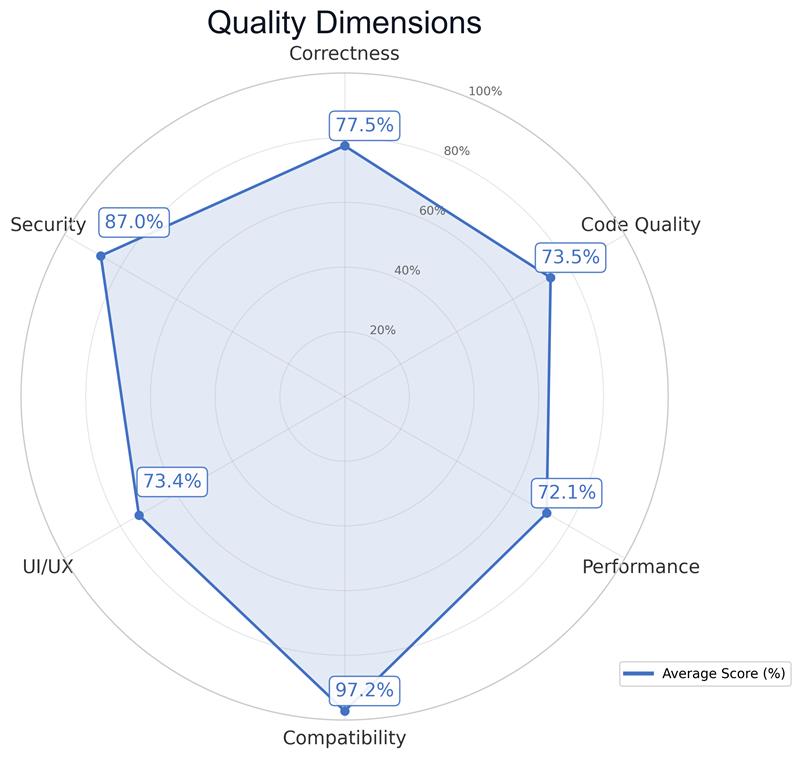}
\caption{Overall performance: high platform compatibility (97.2\%) and security (87.0\%) with solid correctness (73.5\%).}
\label{fig:ai4ui_performance}
\end{figure}
\FloatBarrier
\subsection{User Preference}
A total of 200 expert evaluators participated in the study, providing their judgements on the UI design tasks. Based on these judgements, we derived rankings for AI text to app tools, which are summarized in the overall leaderboard below. Since our application is not currently exposed for public use, it could not be submitted to Design Arena for open community voting. However, we plan to include AI4UI in the Design Arena leaderboard in the near future once public access is enabled. In the interim, the scores for other listed tools were obtained from publicly available benchmark reports, specifically the Design Arena leaderboard dated 12 November 2025, to ensure a fair and comprehensive comparison. Additionally, for a subset of frameworks we performed our own internal replication tests under matched task conditions; those independently collected results were used to align and validate AI4UI's relative positioning against the publicly reported Design Arena standings.

\FloatBarrier
\begin{figure}[!htbp]
\centering
\includegraphics[width=1.0\textwidth]{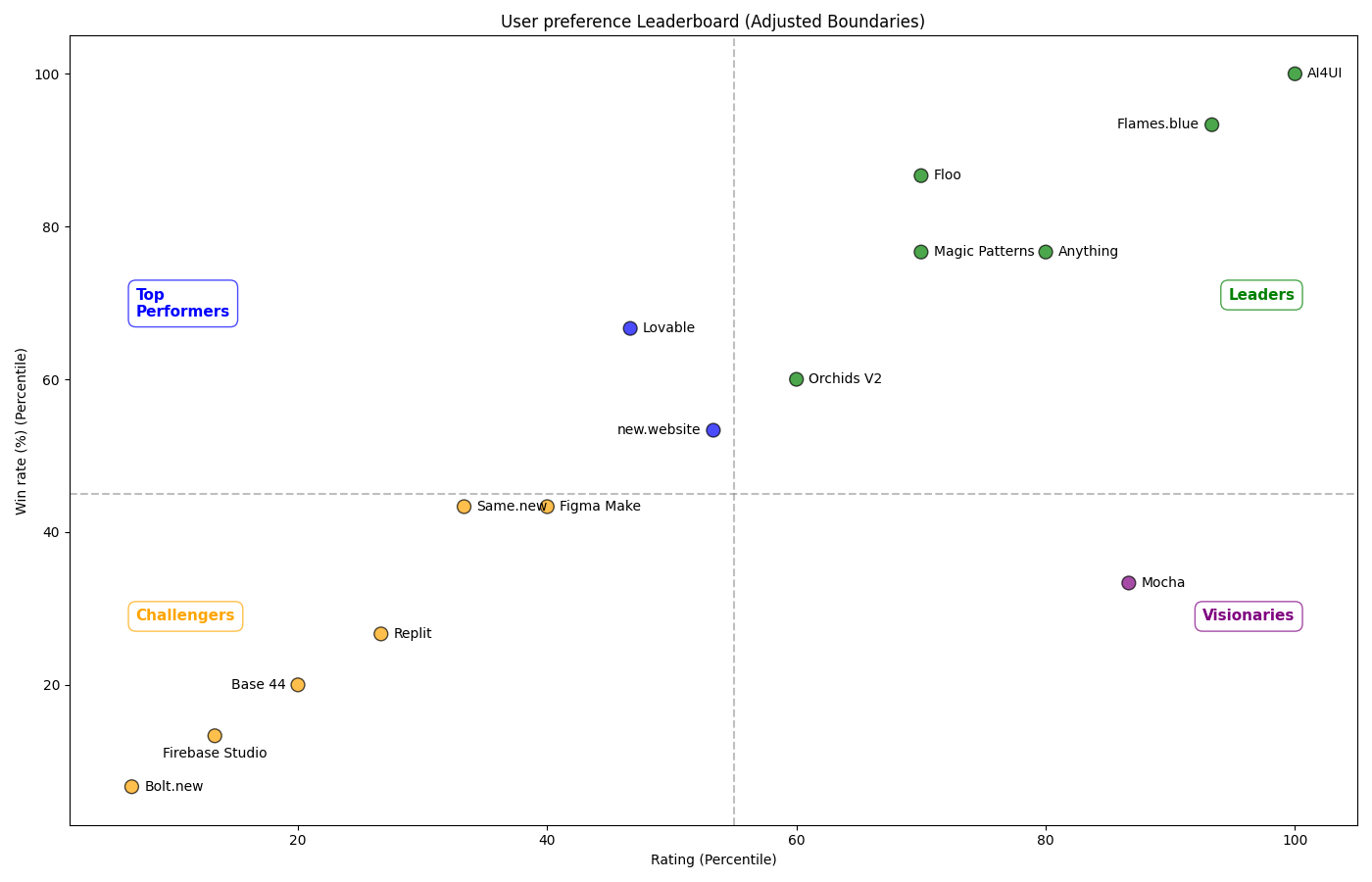}
\caption{User preference leaderboard}
\label{fig:ai4ui_performance}
\end{figure}
\FloatBarrier

A high user preference score signals that the tool not only meets technical requirements but also produces designs that users want to adopt, a critical factor in product acceptance and overall satisfaction~\cite{dvir2018how, jankowski2019gradual}. For enterprise deployments, strong user preference enhances stakeholder buy-in, improves adoption rates, and reinforces the credibility of the AI-assisted development pipeline. It should be noted that the expert evaluators who participated in this study were predominantly from the South-East Asian demographic, which may introduce subtle regional biases in design preferences, aesthetic judgments, and usability expectations. While these preferences align well with our target deployment contexts, broader geographic validation could affect the relative positioning in the rankings.

\FloatBarrier 

\section{The Path Forward}
\label{sec:future}
As industry moves toward AI-augmented enterprise development, specialized autonomous agents will increasingly outperform generalist AI solutions~\cite{srivastava2023beyond, zhong2023agieval}. AI4UI's integrated workflow from design grammar integration and autonomous pipeline execution to targeted human augmentation, empowers organizations to:
\begin{itemize}
\item Accelerate development timelines from months to weeks
\item Maintain enterprise-grade quality and security standards
\item Scale frontend development across multiple teams and projects
\item Reduce technical debt through expertise-driven architecture
\end{itemize}

\section{Conclusion}
\label{sec:conclusion}
AI4UI represents a significant leap in enterprise-grade front-end development, combining autonomous code generation with strategic human-in-the-loop stages that preserve architectural authority and quality. Validated through rigorous benchmarking and expert preference studies, it delivers production-ready UI at scale, rivaling the highest-performing solutions in the field.

What sets AI4UI apart is its focus on engineering readiness, not just visual fidelity~\cite{ngo2003modelling, o'donovan2011color}. By integrating domain-aware knowledge graphs, configuration-driven workflows, and a layered quality assurance pipeline, AI4UI efficiently bridges the gap between design intent and deployable production code.

For organizations accelerating digital transformation, AI4UI is more than a development framework, it is a strategic enabler for autonomous software engineering, offering faster delivery, reduced technical debt, and enterprise-grade robustness.

\section*{Acknowledgments}

This work was conducted as part of the Infosys Finacle platform development initiative. We thank the engineering teams for their contributions to the autonomous agent framework and evaluation studies. We also acknowledge the 200 expert evaluators who participated in the user preference studies.

\section*{Data Availability}

The benchmark results and evaluation metrics presented in this paper are available upon reasonable request. The AI4UI agent framework and associated source code are proprietary components of the Infosys Finacle platform and are not publicly available due to commercial licensing restrictions.

\nocite{*} 
\bibliographystyle{plain}
\bibliography{references}

\end{document}